\documentclass[amsmath,amssymb,12pt,aps]{revtex4-2}
\usepackage[utf8]{inputenc} 
\usepackage{graphicx}
\usepackage{float}
\usepackage{natbib}
\usepackage{wrapfig}
\usepackage{xcolor}
\usepackage{dcolumn}
\usepackage[english]{babel}
\begin{document}
	
\title{On the numerical integration of the multidimensional Kuramoto model}

\author{Marcus A. M. de Aguiar }

\affiliation{Instituto de F\'isica `Gleb Wataghin', Universidade Estadual de Campinas, Unicamp 13083-970, Campinas, SP, Brazil}

\begin{abstract}
	
	The Kuramoto model, describing the synchronization dynamics of coupled oscillators, has been generalized in many ways over the 
	past years. One recent extension of the model replaces the oscillators, originally characterized by a single phase, by particles with $D-1$ 
	internal phases, represented by a point on the surface of the unit D-sphere. Particles are then more easily represented by $D$-dimensional 
	unit vectors than by $D-1$ spherical angles. However,  numerical integration of the state equations should ensure that the propagated vectors 
	remain unit and that particles rotate on the sphere as predicted by the dynamical equations. As discussed in  \cite{lee2023improved} integration 
	of the three-dimensional Kuramoto model using Euler's  method with time step $\Delta t$ not only changes the norm of the vectors but produces 
	a small rotation of the particles around the wrong axis. Importantly, 	the error in the axis' direction does not vanish in the limit $\Delta t \rightarrow 0$. 
	Therefore, instead of displacing the unit vectors in the direction of the velocity one should performed a sequence of direct small rotations, as dictated 
	by the equations of motion. This keeps the particles on the sphere at all times, ensuring 	exact norm preservation, and rotates the particles around 
	the proper axis for small $\Delta t$ \cite{lee2023improved}.  Here I propose an alternative way to do such integration by rotations 
	in 3D that can be generalized to more dimensions using Cayley-Hamilton's theorem. Explicit formulas are provided for 2, 3 and 4 dimensions. 
	I also compare the results with the forth order Runge-Kutta method, which 	seems to provide accurate results even requiring 
	renormalization of the vectors after each integration step.
	
\end{abstract}

\maketitle

\section{Introduction}

The Kuramoto model became a paradigm in the study of  synchronization dynamics \cite{Kuramoto1975,Kuramoto1984}.
It has been used to describe a variety of systems, such as circadian rhythms \cite{yamaguchi2003,bick2020understanding}, power grids \cite{filatrella2008analysis,motter2013spontaneous,Nishikawa_2015,molnar2021asymmetry}, neuronal networks 
\cite{cumin2007generalising,bhowmik2012well,ferrari2015phase,reis2021bursting} and coupled metronomes \cite{Pantaleone2002}.  
The model has also been extended in many ways, with the introduction of frustration \cite{sakaguchi1986soluble,yue2020model,buzanello2022matrix,de2023generalized}, 
different types of coupling functions \cite{hong2011kuramoto,yeung1999time,breakspear2010generative}, networks  \cite{Rodrigues2016,Joyce2019},
distributions of the oscillator's natural frequencies  \cite{Gomez-Gardenes2011,Ji2013}, inertial terms \cite{Acebron2005,dorfler2011critical,olmi2014hysteretic}, external periodic driving forces
\cite{Childs2008,moreira2019global,moreira2019modular} and coupling with particle swarms \cite{o2017oscillators,o2022collective,supekar2023learning}.

The original model describes the dynamics of $N$ oscillators, represented by their phases $\theta_i$, and coupled according to the equations 
\begin{equation}
	\dot{\theta}_i = \omega_i + \frac{k}{N} \sum_{j=1}^N \sin{(\theta_j-\theta_i)}
	\label{kuramoto}
\end{equation}
where  $\omega_i$ are their natural frequencies, selected from a symmetric distribution $g(\omega)$, $k$ is the coupling strength and $i = 1, ..., N$. Kuramoto showed that,
for $k$ is sufficiently large, the oscillators synchronize their phases, behaving as a single particle. The transition to
synchronization can be described by the complex order parameter 
\begin{equation}
	z = p e^{i \psi} \equiv \frac{1}{N} \sum_{i=1}^N e^{i\theta_i}
	\label{paraord}
\end{equation}
with disordered motion resulting in $p \approx 0$ and coherent motion in $p \approx 1$. In the limit $N \rightarrow \infty$, the onset of synchronization can 
be described as a continuous phase transition, where $p=0$ for $k < k_c =  2/\pi g(0)$ and increases as $p = \sqrt{1-k_c/k}$ for $ k > k_c$ \cite{Acebron2005,Rodrigues2016}. 

Recently, Chandra et al \cite{chandra2019continuous} have shown that Kuramoto oscillators could also be represented by unit vectors $\vec{\sigma_i} = (\cos{\theta_i},\sin{\theta_i})$ 
rotating on the unit circle. According to Eq.(\ref{kuramoto}),  the dynamics of $\vec{\sigma}_i$ is given by 
\begin{equation}
	\frac{d \vec{\sigma_i}}{d t} = \mathbf{W}_i \vec{\sigma_i} + \frac{k}{N} \sum_j [\vec{\sigma_j} - (\vec{\sigma_i}\cdot \vec{\sigma_j}) \vec{\sigma_i}]
	\label{eq3}
\end{equation}
where  $\mathbf{W}_i$ is the anti-symmetric matrix 
\begin{equation}
	\mathbf{W}_i = \left( 
	\begin{array}{cc}
		0 & -\omega_i \\
		\omega_i & 0
	\end{array}
	\right).
	\label{wmat}
\end{equation}
The complex order parameter  $z$, Eq.(\ref{paraord}), can be written in terms of the real vector
\begin{equation}
	\vec{p} = \frac{1}{N}\sum_i \vec{\sigma_i} = (p\cos\psi,p\sin\psi)
	\label{vecpar}
\end{equation}
describing the center of mass of the system. 

\section{Multidimensional Kuramoto model}

Eq.(\ref{eq3}) can be extended to higher dimensions by simply considering unit vectors $\vec\sigma_i$ in D-dimensions, rotating on the surface of the corresponding (D-1) unit sphere. The matrices $\mathbf{W}_i$ become $D \times D$ anti-symmetric matrices containing the $D(D-1)/2$ natural frequencies of each oscillator. Finally, the $D$-dimensional model is further extended by replacing the coupling constant $k$ by a coupling matrix $\mathbf{K}$ \cite{barioni2021complexity,buzanello2022matrix,de2023generalized}:
\begin{equation}
	\begin{array}{ll}
		\displaystyle{
	\frac{d \vec{\sigma_i}}{d t} }&= \mathbf{W}_i \vec{\sigma_i} + \frac{1}{N} \sum_j [{\mathbf K} \vec{\sigma_j} - (\vec{\sigma_i}\cdot {\mathbf K} \vec{\sigma_j}) \vec{\sigma_i}]  \\ \\
	  &=  \mathbf{W}_i \vec{\sigma_i} +  [ \mathbf{K} \vec{p}- (\vec{\sigma_i}\cdot \mathbf{K} \vec{p}) \vec{\sigma_i}].
\end{array}
	\label{kuragen}
\end{equation}
The coupling matrix breaks the rotational symmetry and plays the role of a generalized frustration: it rotates $\vec\sigma_j$, hindering its alignment with $\vec\sigma_i$ and inhibiting synchronization.  The angle of rotation depends on $\sigma_j$, generalizing the constant frustration angle of the Sakaguchi model \cite{sakaguchi1986soluble}.  Norm conservation, $|\vec{\sigma_i}|=1$, is guaranteed, as can be seen by taking the scalar product of Eqs.(\ref{kuragen}) with $\vec{\sigma_i}$.  Similar extensions of the Kuramoto model to higher dimensions were also considered in  refs. \cite{Tanaka2014,lipton2021kuramoto,crnkic2021synchronization}.

\section{Dynamics on the sphere}

Eq.(\ref{kuragen}) describes the dynamics of particles on the surface of a $D$-dimensional unit sphere and the right hand side, expressing
the velocity of the i-th oscillator, is tangent to the sphere at all times. Euler's method attempts to solve these equations by 
considering the approximation where $\vec{\sigma}_i(t+\Delta t) \approx \vec{\sigma}_i(t) + \dot{\vec{\sigma}}_i(t) \Delta t$. This,
however, moves the particle away from the sphere and increases the norm of $\vec{\sigma}_i$, requiring it to be manually renormalized as
\begin{equation}
	\vec{\sigma}_i(t+\Delta t) \approx \frac{ \vec{\sigma}_i(t) + \dot{\vec{\sigma}}_i(t) \Delta t}{|\vec{\sigma}_i(t) + \dot{\vec{\sigma}}_i(t) \Delta t|}.
\end{equation}
Although the error in the norm is of order $\Delta t^2$, it was shown in \cite{lee2023improved} that Euler's procedure introduces an error in the 
particle's axis of rotation that does not vanish in the limit $\Delta t \rightarrow 0$. In order to overcome this problem, a new integration 
method, relying on a sequence of rotations using Euler's angles, was proposed and shown to converge to correct
result as $\Delta t \rightarrow 0$ \cite{lee2023improved}.

Here I will follow the ideas introduced in \cite{lee2023improved} and propose a slightly different integration scheme that also preserves norm 
exactly and that can be extended to higher dimensions. In $D=3$ the method is equivalent to that proposed in \cite{lee2023improved}, but easier to
implement numerically. I start by defining
\begin{equation}
	\vec{p}_k = \mathbf{K} \vec{p}
	\label{pk}
\end{equation}
and noting that the second term on the RHS of  Eq. (\ref{kuragen}) can be written as 
\begin{equation}
	  \mathbf{K} \vec{p}- (\vec{\sigma_i} \cdot \mathbf{K} \vec{p}) \vec{\sigma_i} =  \vec{p}_K- (\vec{\sigma_i} \cdot \vec{p}_K) \vec{\sigma_i}  = 
	   \mathbf{U}_i \vec{\sigma}_i
\end{equation}
where 
\begin{equation}
	\mathbf{U}_i =  \vec{p}_k \, \vec{\sigma_i}^T -  \vec{\sigma_i} \, \vec{p}_k^{\,T} 
	\label{basic}
\end{equation}
is an anti-symmetric matrix. The superscript $T$ stands for transpose and the dyadic matrices are defined by $(\vec{A} \, \vec{B}^T)_{ij} = A_iB_j$ and satisfy $(\vec{A} \vec{B}^T) \vec{C} = (\vec{B} \cdot \vec{C}) \vec{A} $.  Equation (\ref{kuragen}) can now be written as
\begin{equation}
	\frac{d \vec{\sigma_i}}{d t} = \mathbf{V}_i \vec{\sigma_i} 
	\label{kuramotogenkh}
\end{equation}
where 
\begin{equation}
	\mathbf{V}_i = \mathbf{W}_i + \mathbf{U}_i = \mathbf{W}_i  + \vec{p}_k \, \vec{\sigma_i}^T -  \vec{\sigma_i} \, \vec{p}_k^{\,T} 
\end{equation}
is itself anti-symmetric. 

If the matrix $\mathbf{V}_i$ were time-independent, the solution of Eq.(\ref{kuramotogenkh}) would be $\sigma_i(t) = e^{\mathbf{V}_i t} \vec{\sigma_i}(0)$. 
However, because  $\mathbf{V}_i$ depends not only on the time but also on the full set of particles' positions $\vec{\sigma_j}$, this is not true. Still, one
can approximate
\begin{equation}
	\vec{\sigma_i}(t+\Delta t) \approx e^{\mathbf{V}_i (t) \Delta t} \vec{\sigma_i}(t)
	\label{evol}
\end{equation}
where $e^{\mathbf{V}_i (t) \Delta t}$ plays the role of an infinitesimal evolution operator. The
error in this approximation comes only from the fact that $\mathbf{V}_i$ does not remain constant throughout the time interval from $t$ to $t+\Delta t$.
Norm conservation, though, is exact: since $\mathbf{V}_i^T = - \mathbf{V}_i$, $|\vec{\sigma}_i(t+\Delta t)|^2 = \vec{\sigma}_i^T(t) e^{\mathbf{V}_i^T(t)} e^{\mathbf{V}_i (t)}  \vec{\sigma_i}(t) = |\vec{\sigma_i}(t)|^2$.

In three dimensions Eq.(\ref{kuramotogenkh})  is equivalent to 
\begin{equation}
	\frac{d \vec{\sigma_i}}{d t} = \vec{\Omega}_i \times \vec{\sigma_i} 
	\label{kuramotogenkho}
\end{equation}
where the components of $\vec{\Omega}_i$ are related to $\mathbf{V}_i$ by 
\begin{equation}
	\mathbf{V}_i = \left( 
	\begin{array}{ccc}
		0 & -\Omega_{i3} & \Omega_{2i} \\
		\Omega_{i3} & 0  & -\Omega_{i1} \\
		-\Omega_{i2} & \Omega_{i1} & 0
	\end{array}
	\right)
	\label{vmat3}
\end{equation}
with
\begin{equation}
	\begin{array}{lll}
		\Omega_{i1} & = & \omega_{i1} + p_{k2} \sigma_{i3} - p_{k3} \sigma_{i2} \\
		\Omega_{i2} & = & \omega_{i2} + p_{k3} \sigma_{i1} - p_{k1} \sigma_{i3} \\
		\Omega_{i3} & = & \omega_{i3} + p_{k1} \sigma_{i2} - p_{k2} \sigma_{i1} .
	\end{array}
	\label{vcomp3}
\end{equation}

Equation (\ref{kuramotogenkho}) explicitly represents  a rotation on the sphere. The vector $\hat{\Omega}_i$ gives the instantaneous axis of rotation, and depends 
on the natural frequencies of particle $i$ and on the position of all other oscillators through $\vec{p}$. This equation can be used directly to 
integrate the dynamics, avoiding the artifacts introduced by 
changing the norm of the unit vectors $\vec{\sigma}_i$ and projecting them back to the sphere \cite{lee2023improved}. However, this form of vector 
product does not hold in more dimensions. In the next sections I will propose an integration 
method, based on Eq.(\ref{kuramotogenkh}), instead of Eq.(\ref{kuramotogenkho}), that is perhaps simpler in three dimensions and that generalizes to 
higher dimensions as well.  I shall give explicit formulas  for $D=2$, $D=3$ and $D=4$.

\section{Three Dimensions}

Using Cayley-Hamilton's theorem we write 
\begin{equation}
	e^{\mathbf{V}_i \Delta t} =  \sum_{k=0}^{D-1} \beta_k  \mathbf{V}_i^k 
	\label{evd}
\end{equation}  
where $D$ is the dimension of $\mathbf{V}_i$. This expression also holds for the eigenvalues $\lambda$ of $\mathbf{V}_i$,
\begin{equation}
	e^{\lambda \Delta t} =  \sum_{k=0}^{D-1} \beta_k  \lambda^k.
	\label{evda}
\end{equation}  
Writing Eq.(\ref{evda}) for the $D$ eigenvalues of $\mathbf{V}_i$ gives a linear system that can be solved for the coefficients $\beta_k$. For $D=3$ the
eigenvalues are $\pm i \Omega_i$ and $0$, which results in 
\begin{equation}
	e^{\mathbf{V}_i \Delta t} = \mathbf{1} + \frac{\sin(\Omega_i \Delta t)}{\Omega_i}  \mathbf{V}_i + 
		\frac{1 - \cos(\Omega_i \Delta t)}{\Omega_i^2}  \mathbf{V}_i^2.
\end{equation}
Computing $\mathbf{V}_i^2$ I find
\begin{equation}
	\mathbf{V}_i^2 = \vec{\Omega}_i \vec{\Omega}_i^T - \Omega_i^2 \mathbf{1} 
	\label{v23d}
\end{equation}
where
\begin{equation}
\Omega_i^2 = \Omega_{i1}^2 + \Omega_{i2}^2 + \Omega_{i3}^2,
\end{equation}
and obtain 
\begin{equation}
	e^{\mathbf{V}_i \Delta t} = \cos(\Omega_i \Delta t) \mathbf{1} + \frac{\sin(\Omega_i \Delta t)}{\Omega_i}  \mathbf{V}_i + 
	\frac{1 - \cos(\Omega_i \Delta t)}{\Omega_i^2}   \vec{\Omega}_i \vec{\Omega}_i^T.
\end{equation}
Finally, Eq.(\ref{evol}) becomes 
\begin{equation}
	\vec{\sigma_i}(t+\Delta t) = \cos(\Omega_i \Delta t) \vec{\sigma}_i + \sin(\Omega_i \Delta t) \hat{\Omega}_i \times \vec{\sigma}_i + 
	(1 - \cos(\Omega_i \Delta t))   (\hat{\Omega}_i \cdot \vec{\sigma}_i) \hat{\Omega}_i
	\label{evol3d}
\end{equation}
where $\hat{\Omega}_i = \vec{\Omega}_i/\Omega_i$ and all quantities on the RHS are computed at time $t$. Eq.(\ref{evol3d}) corresponds
to a rotation of $\vec{\sigma}_i$ around $\hat{\Omega}_i$ by the angle $\Omega_i \Delta t$ \cite{goldstein2002classical}. Once all vectors 
$\vec{\sigma}$ have been updated, $\vec{p}_k$ and $\vec{\Omega}_i$'s are recalculated and the process iterated. Eq. (\ref{evol3d}) 
contains all orders of the exponential and would be exact if $\vec{\Omega}_i$ were  constant. The error of the integration process 
can be estimated by calculating the how fast the matrix elements of $\mathbf{U}_i$, responsible for 
the interactions, are changing over time. This allows us to adjust the time step $\Delta t$ according its rate of change, increasing it when the rate is 
smaller than a specified threshold and reducing it otherwise. This procedure can reduce the integration time considerably.

\section{Four dimensions}

In $D=4$ equations (\ref{basic}) and (\ref{kuramotogenkh}) still work but the exponential of $\mathbf{V}_i = \mathbf{W}_i + \mathbf{U}_i$
is more complicated for three reasons: first, $\mathbf{W}_i$ has six independent components; second, the eigenvalues of $\mathbf{V}_i$ are
not immediately available and; third, the exponential of $\mathbf{V}_i$ involves powers $\mathbf{V}_i^2$ and $\mathbf{V}_i^3$, which need
to be calculated. Using Cayley-Hamilton's formula we write
\begin{equation}
	e^{\mathbf{V}_i \Delta t} = \beta_0 \mathbf{1} + \beta_1 \mathbf{V}_i + \beta_2 \mathbf{V}_i ^2 + \beta_3 \mathbf{V}_i^3 
	\label{ev4}
\end{equation}  
In the case of identical oscillators, $\mathbf{W}_i=0$, Eq.(\ref{basic}) can be used to write down the powers of $\mathbf{V}_i$ in terms of dyadic matrices as
\begin{equation}
	\mathbf{V}_i^2 = (\vec{p}_k \cdot \vec{\sigma}_i) [\vec{p}_k \, \vec{\sigma}_i ^{\, T} + \vec{\sigma}_i \, \vec{p}_k^{\, T}] - \vec{p}_k \, \vec{p}_k^{\, T} - p_k^2 \vec{\sigma}_i \, \vec{\sigma}_i^{\, T}
\end{equation}
and 
\begin{equation}
	\mathbf{V}_i^3 = [(\vec{p}_k \cdot \vec{\sigma}_i)^2-p_k^2] [\vec{p}_k \, \vec{\sigma}_i ^{\, T} - \vec{\sigma}_i \, \vec{p}_k^{\, T}] .
\end{equation}
In the general case, however, there is no simple form for $\mathbf{V}_i^2$ or $\mathbf{V}_i^3$ like in the 3D case (see Eq. (\ref{v23d})) and they
have to be computed numerically as shown below.

To compute the coefficients $\beta_k$ in Eq.(\ref{ev4}) we need the eigenvalues of $\mathbf{V}_i$. For each eigenvalue $\lambda$ it holds that
\begin{equation}
	e^{\lambda \Delta t} = \beta_0  + \beta_1 \lambda+ \beta_2 \lambda ^2 + \beta_3 \lambda^3.
	\label{ev4a}
\end{equation}  
Writing this equation for the four eigenvalues allows us to find all $\beta$'s. Because $\mathbf{V}_i$ is 
anti-symmetric, its eigenvalues come in pairs of complex conjugates, $\pm i \alpha_1$ and $\pm i \alpha_2$. The result is
\begin{equation}
	\begin{array}{ll}
		\beta_0 &= [\alpha_1^2 \cos(\alpha_2 \Delta t)  - \alpha_2^2   \cos(\alpha_1 \Delta t)]/(\alpha_1^2 - \alpha_2^2)\\
		\beta_1 &=  [\alpha_1^3 \sin(\alpha_2 \Delta t)  - \alpha_2^3   \sin(\alpha_1 \Delta t)]/[(\alpha_1 \alpha_2)(\alpha_1^2 - \alpha_2^2)]\\
		\beta_2 &= [\cos(\alpha_2 \Delta t)  - \cos(\alpha_1 \Delta t)]/(\alpha_1^2 - \alpha_2^2)\\
		\beta_3 &=  [\alpha_1 \sin(\alpha_2 \Delta t)  - \alpha_2   \sin(\alpha_1 \Delta t)]/[(\alpha_1 \alpha_2)(\alpha_1^2 - \alpha_2^2)].
	\end{array}
	\label{beta}
\end{equation}

Finally, writing $\mathbf{V}$ explicitly as
\begin{equation}
	\mathbf{V} = \left(
	\begin{array}{cccc}
		0 & a & b & c \\
		-a & 0 & d & e \\
		-b & -d& 0 &f \\
		-c & -e & -f &0
	\end{array}
	\right)
\end{equation}
I obtain
\begin{equation}
	\begin{array}{ll}
		\alpha_1 &= \sqrt{\frac{\Omega^2}{2} + \frac{1}{2} \sqrt{\Omega^2 - 4 \Lambda}} \\ \\
		\alpha_2 &= \sqrt{\frac{\Omega^2}{2} - \frac{1}{2} \sqrt{\Omega^2 - 4 \Lambda}} 
	\end{array}
	\label{ang1}
\end{equation}
where
\begin{equation}
	\begin{array}{ll}
		\Omega^2 &= a^2 + b^2 + c^2 + d^2 + e^2 + f^2\\
		\Lambda &= a^2f^2 + e^2b^2 + c^2 d^2 - 2afeb +2afcd-2abcd.
	\end{array}
	\label{ang2}
\end{equation}

The algorithm for integrating the equations in $D=4$ can then be summarized as follows: for the initial distribution of
oscillators, compute $\vec{p}$ and $\vec{p}_k = \mathbf{K} \vec{p}$. For each oscillator $i$ compute  $\mathbf{V}_i = 
\mathbf{W}_i +  \vec{p}_k \, \vec{\sigma_i}^T -  \vec{\sigma_i} \, \vec{p}_k^{\,T}$ and use Eqs. (\ref{ang1}) and (\ref{ang2})
to compute the angles $\alpha_1$ and $\alpha_2$. Compute the coefficients $\beta_k$ using Eq.(\ref{beta}) and update
the oscillator's position using 
\begin{equation}
	\vec{\sigma}_i (t+\Delta t) = \beta_0 \vec{\sigma}_i(t) + \beta_1 \mathbf{V}_i \vec{\sigma}_i(t) + 
	\beta_2 \mathbf{V}_i ^2 \vec{\sigma}_i(t) + \beta_3 \mathbf{V}_i^3 \vec{\sigma}_i(t).
	\label{evol4d}
\end{equation}

\section{higher dimensions and D=2}

The formal evolution equations provided by Eqs.(\ref{evol3d}) and (\ref{evol4d}) extends naturally to any number $D$ of dimensions. In $D=5$ one of the eigenvalues of $\mathbf{V}_i$ is always zero, implying $\beta_0=1$. The remaining four  eigenvalues can be obtained analytically with expressions similar to Eqs. (\ref{ang1})  and (\ref{ang2}). In 6 or more dimensions, however, solutions involve cubic or higher order equations
and need to be performed numerically. 

The two dimensional case, corresponding to the Kuramoto model and its generalizations in 2D \cite{buzanello2022matrix}, can be easily integrated using the same formalism. In this case 
\begin{equation}
	\mathbf{V}_i = \left(
	\begin{array}{cc}
		0 & v_i \\
		-v_i & 0
	\end{array}
	\right)
\end{equation}
with $v_i = -\omega_i + p_{k1} \sigma_{i2} - p_{k2} \sigma_{i1}$. The evolution operator becomes $e^{\mathbf{V}_i \Delta t} = \cos(v_i \Delta t) + (\sin(v_i \Delta t)/v_i) \mathbf{V}_i = R(v_i \Delta t)$ where $R$ is the usual $2\times 2$ rotation matrix
\begin{equation}
R (\theta)=	\left(
	\begin{array}{ll}
		\cos\theta &  \sin\theta \\
		-  \sin\theta &  \cos\theta 
	\end{array}
	\right)
\end{equation}
as expected. The components of $\vec{\sigma}$ are then easily updated according to $\vec{\sigma} (t+\Delta t) = R(v_i \Delta t) \, \vec{\sigma}(t)$.
 In the next section I will show examples of numerical simulations in $D=3$ and $D=4$.

\section{Simulations}

\subsection{Three Dimensional systems}

\begin{figure}
	\includegraphics[scale=0.29]{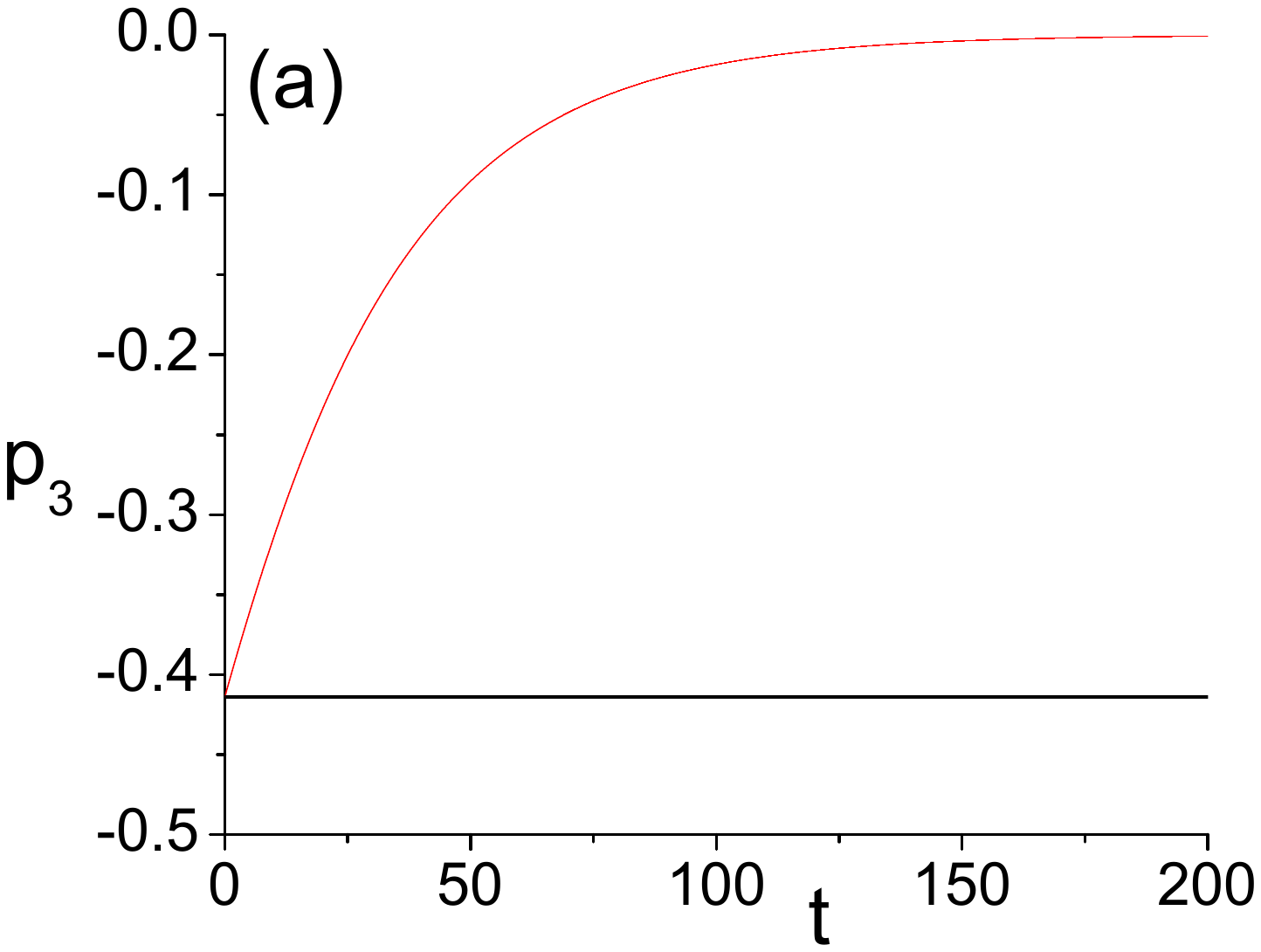} 
	\includegraphics[scale=0.29]{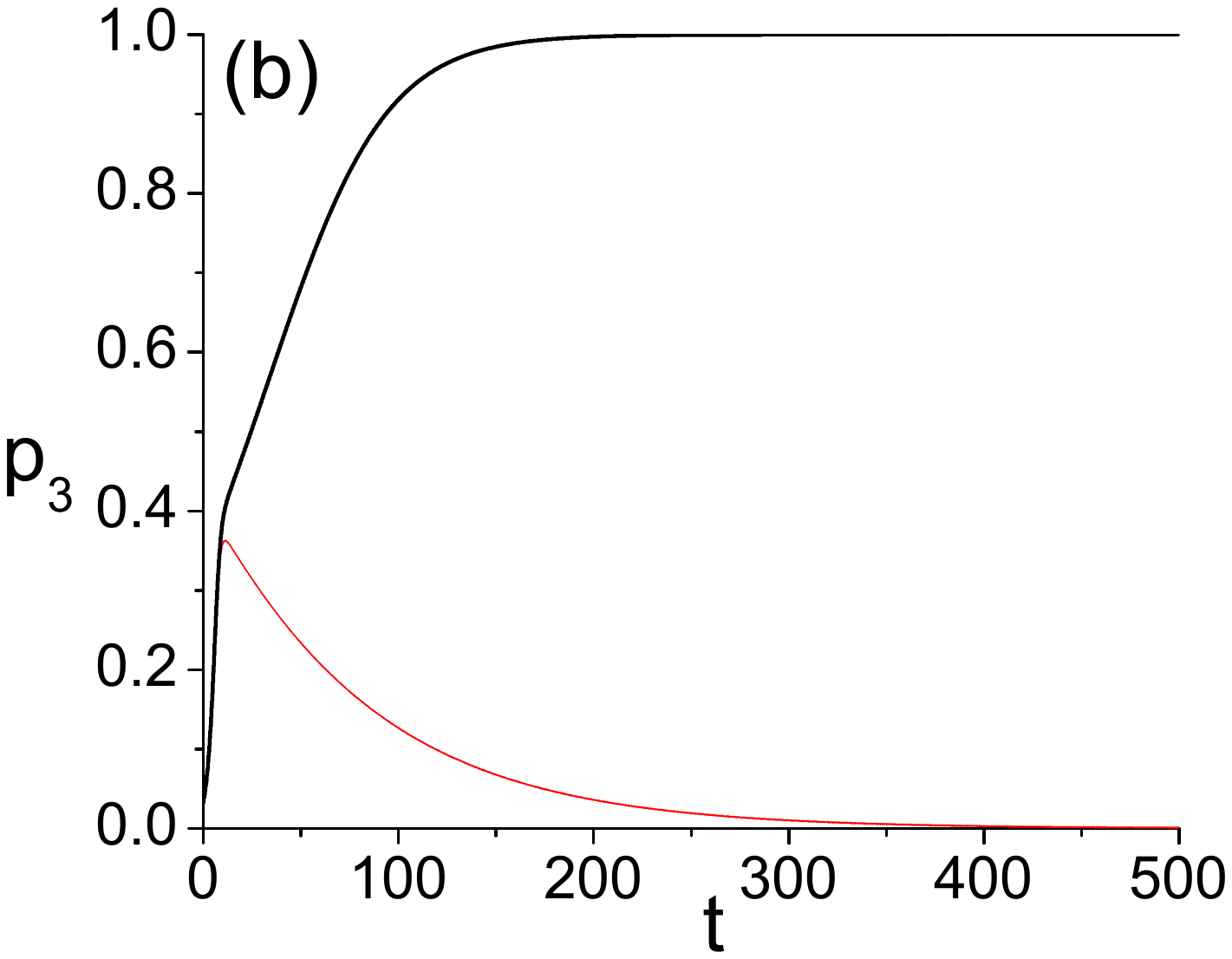} 
	\includegraphics[scale=0.29]{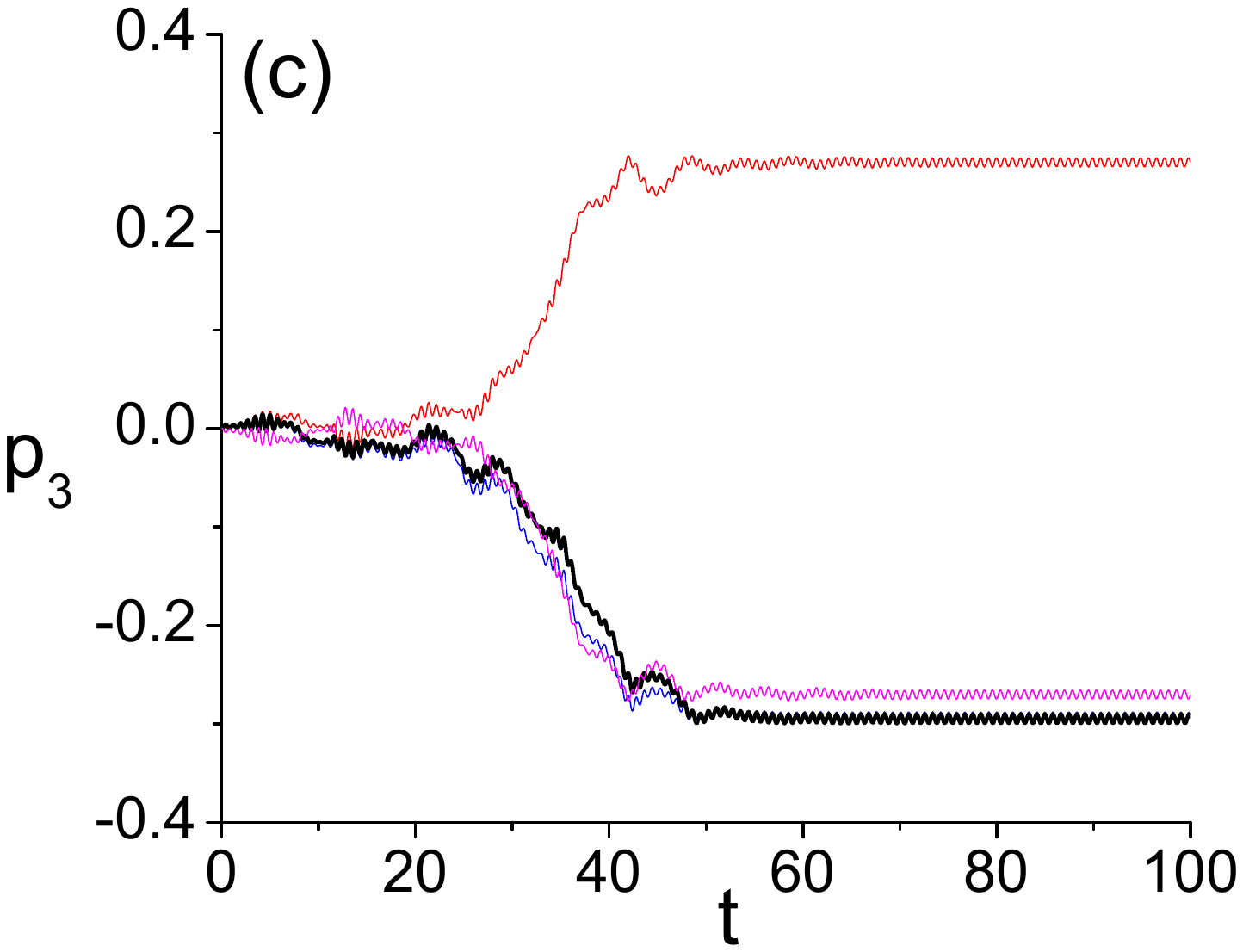} 
	\includegraphics[scale=0.29]{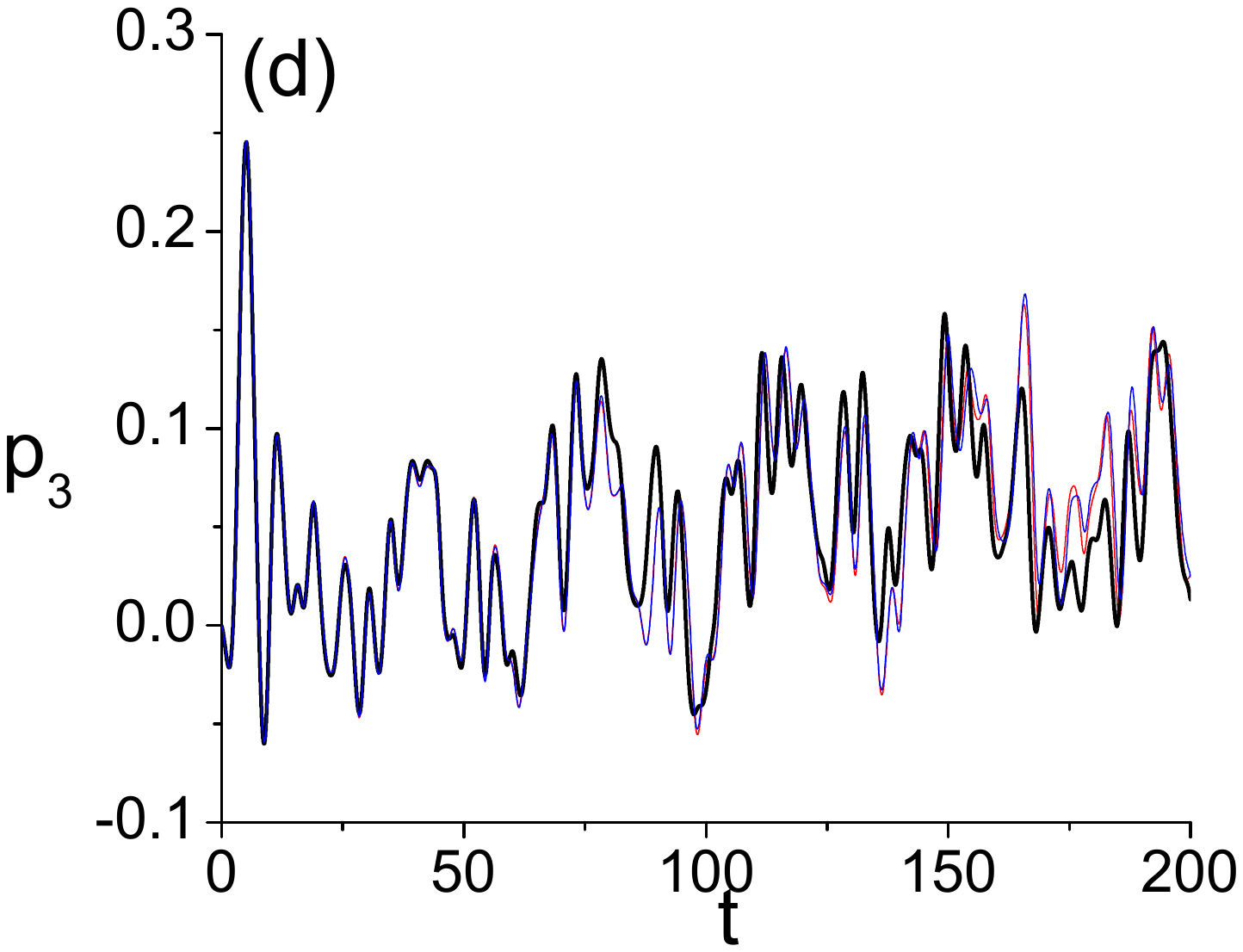} 
	\caption{Third component of order parameter as a function of time according to Euler's method (red thin line), Eq.(\ref{evol3d})(black thick line)
		and 4th order Runge-Kutta (blue thin black line). In all panels the average value of the natural frequencies
		is zero in directions 1 and 2. The coupling matrix is diagonal in panels (a)-(b) with coupling intensity
		0.5. For panels (c)-(d) the coupling matrix is given by Eq.(\ref{cm3}): (a) single particle with $\omega_3=8$; (b) $n=100$ particles
		with $ \langle \omega_3 \rangle =8$; (c) $n=100$ particles	with $ \langle \omega_3 \rangle =8$ (purple line shows $-p_3$ for Euler's method) ; 
		(d) $n=100$ particles	with $ \langle \omega_3 \rangle =0$. }
	\label{fig1}
\end{figure}

We first consider a single oscillator in 3D. In this case $\mathbf{V}$ is time independent and Eq.(\ref{evol3d}) is exact
for any $\Delta t$. Choosing $\vec{\Omega} = (0,0,\omega)$ and $\vec{\sigma}(0) =(\sigma_1,\sigma_2,\sigma_3)$ we can write 
\begin{equation}
	\vec{\sigma}(t) = \cos(\omega t) \vec{\sigma}(0) + \frac{\sin(\omega t)}{\omega}  \vec{\omega} \times \vec{\sigma}(0) + 
	\frac{1 - \cos(\omega t)}{\omega^2}   (\vec{\omega} \cdot \vec{\sigma}(0)) \vec{\omega}
\end{equation}
which results in the obvious rotation $\vec{\sigma}(t) = (\sigma_1 \cos(\omega t) - \sigma_2 \sin(\omega t), \sigma_2 \cos(\omega t) + \sigma_1 \sin(\omega t), \sigma_3 )$. 
Although trivial, this is exactly one of the cases where Euler's method fails \cite{lee2023improved}.  Figure \ref{fig1}(a) shows the third component of 
$\vec{\sigma}(t)$ as computed with Euler's method using a time step of $0.0005$ and compared with the exact constant result $\sigma_3$. 
AS discussed in \cite{lee2023improved}, the reason $\sigma_3(t)$ decreases to zero is that the module of $\vec{\sigma}(t+\Delta t)$ is always slightly 
larger than 1 in Euler's method, as the 
velocity vector is tangent to the sphere. When $\vec{\sigma}(t+\Delta t)$ gets renormalized, by diving it by its module, the third component 
(that should stay constant) decreases a little. Interestingly, integration with 4th order Runge-Kutta method (4RK) agrees with the exact result. I
hypothesize that the accuracy of 4RK comes from the fact that the calculation with time step $\Delta t$ uses intermediate points at $\Delta t/4$, 
$\Delta t/2$ and $3 \Delta t/4$, adjusting the velocity field much better around the sphere. Indeed one can verify that the module of  $\vec{\sigma}(t+\Delta t)$ 
is not  always larger than 1 in RK4 simulations.

As a second example I consider $N=100$ particles with coupling matrix $\mathbf{K} = 0.5 \mathbf{1}$ and  natural frequencies $\mathbf{W}_i$
given by vectors $\vec{\omega}_i = (\omega_{i1}, \omega_{i2}, \omega_{i3} + \bar{\omega})$. The frequencies $\omega_{ij}$ 
are Gaussian distributed with zero average and width $\delta=0.1$, but the particles have an overall tendency to rotate
around $z$ with angular velocity $\bar{\omega}=8$. Initial conditions are chosen randomly over the sphere. 
Fig.\ref{fig1}(b) shows the same integration artifact observed for a single particle.

Finally I consider an example where the coupling matrix $\mathbf{K}$ is not proportional to the identity. I set
\begin{equation}
	\mathbf{K} = \left(
	\begin{array}{ccc}
		k_1 \cos \gamma & \, k_1 \sin \gamma & 0  \\
		-k_1 \sin \gamma  & k_1 \cos\gamma & 0  \\
		 0 & 0 & k_2 \\
	\end{array}
	\right)
	\label{cm3}
\end{equation}
with $k_1=1$, $k_2=0.5$ and $\gamma=0.5$. As demonstrated in \cite{de2023generalized} this would result in a synchronized
state where the order parameter rotates in the x-y plane if all the averages natural frequencies were zero.  Fig. \ref{fig1}(c) shows results for the same
settings as in fig. \ref{fig1}(b), i.e., $\omega_{ij}$ Gaussian distributed with zero average, width $\delta=0.1$ and $\bar{\omega_3}=8$.
We see that the third component of $\vec{p}$ increases from near zero, but Euler's method imprecision makes it go in the wrong direction.
Even correcting the sign of $p_3$, its module does not converge to the correct value. Panel (d) shows a similar simulation with $\bar{\omega_3}=0$.
Now the error is Euler's method is much smaller, although it is still there and propagate for longer times.

\subsection{Four Dimensional Systems}

\begin{figure}
	\includegraphics[scale=0.29]{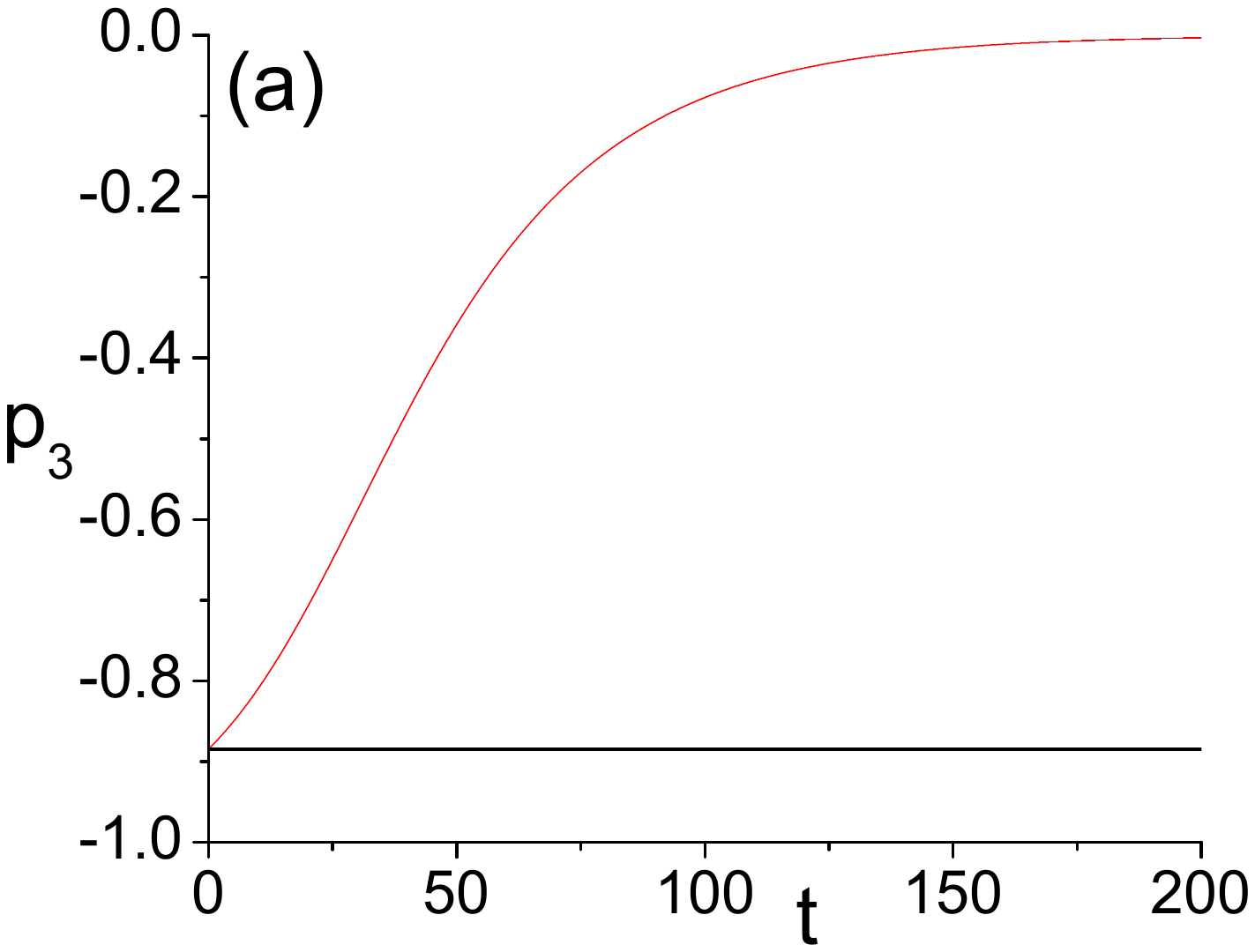} 
	\includegraphics[scale=0.29]{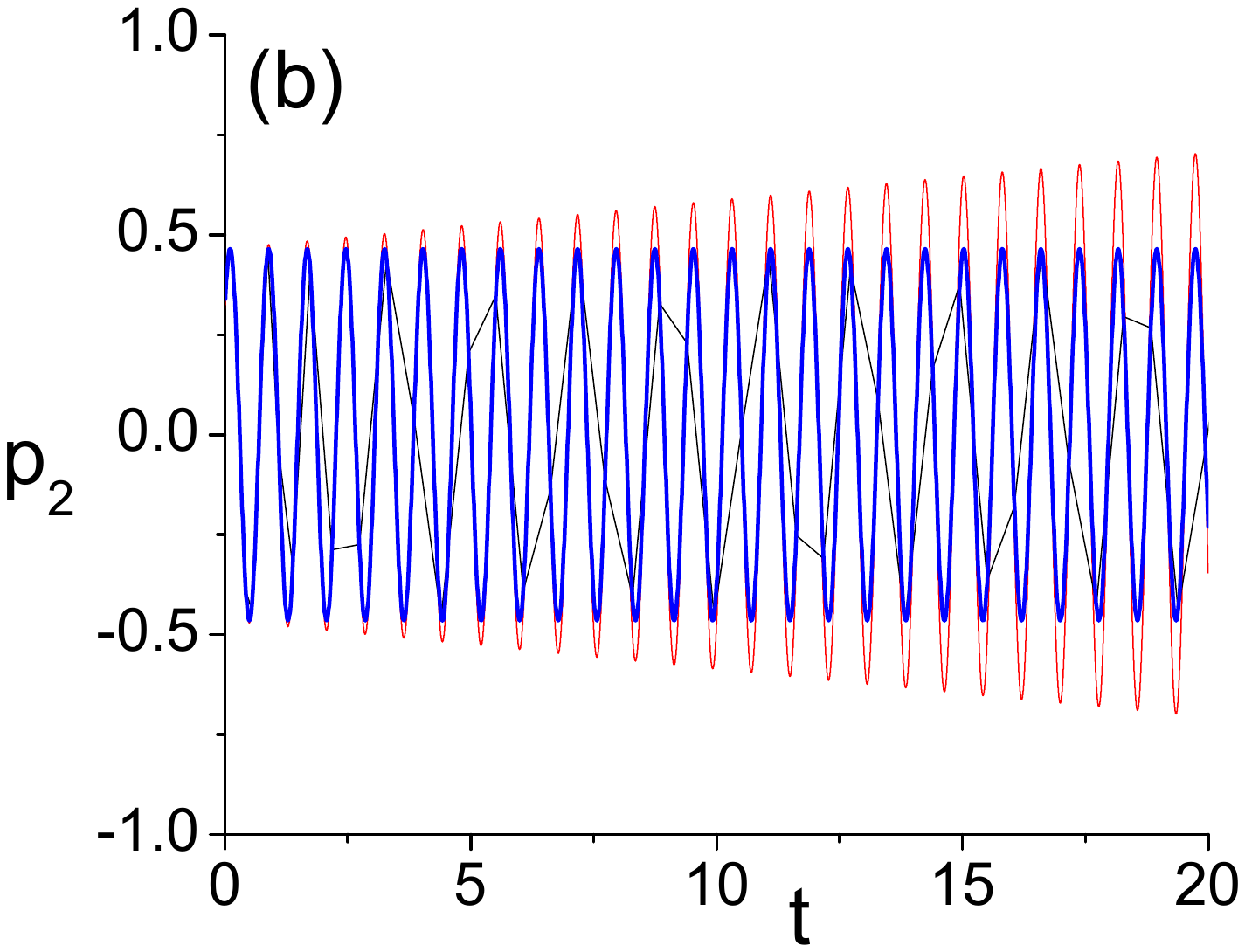} 
	\caption{Time evolution of the (a) third component and (b) second component of the order parameter for a single particle with $\omega=8$. Red 
		thin line shows the result according to Euler's method (red thin line). Evolution according to Eq.(\ref{evol4d}) is shown as blue thick line (using fixed 
		time step 0.01) and black thin line (with adjustable time step). The coupling matrix is diagonal with coupling intensity 3. }
	\label{fig2}
\end{figure}

Again I consider a single particle with coupling matrix proportional to the identity as a first example. In this case $\mathbf{U}=0$ I choose $\mathbf{W}$ as
\begin{equation}
	\mathbf{V}= \mathbf{W} = \left(
	\begin{array}{cccc}
		0 & -\omega & 0 & 0 \\
		\omega & 0 & 0 & 0 \\
		0 & 0 & 0 & 0 \\
		0 & 0 & 0 & 0
	\end{array}
	\right).
\end{equation}
We find $\alpha_1 = \omega$ and $\alpha_2=0$. In this case the coefficients $\beta_k$ simplify to 
\begin{equation}
	\begin{array}{ll}
		\beta_0 &= 1 \\
		\beta_1 &=  \Delta t \\
		\beta_2 &= [1 - \cos(\omega \Delta t)]/ \omega^2\\
		\beta_3 &=  [\omega \Delta t - \sin(\omega \Delta t)]/ \omega^3.\\
	\end{array}
	\label{beta1}
\end{equation}
The matrices $\mathbf{V}^2$ and $\mathbf{V}^3$ can be readily calculated and the results is 
$\vec{\sigma}(t) = (\sigma_1 \cos(\omega t) - \sigma_2 \sin(\omega t), \sigma_2 \cos(\omega t) + \sigma_1 \sin(\omega t), \sigma_3, \sigma_4 )$.
Fig.\ref{fig2} shows $p_3$ and $p_2$, the third and second components of order parameter $\vec{p}$, as a function of time. Once again
Euler's method causes $p_3$ to decrease, consequently increasing the amplitude of oscillations in $p_2$ and $p_1$. Using Eq.(\ref{evol4d}) with
adjustable time step also gives excellent results, and so does RK4 (not shown).

\begin{figure}
	\includegraphics[scale=0.29]{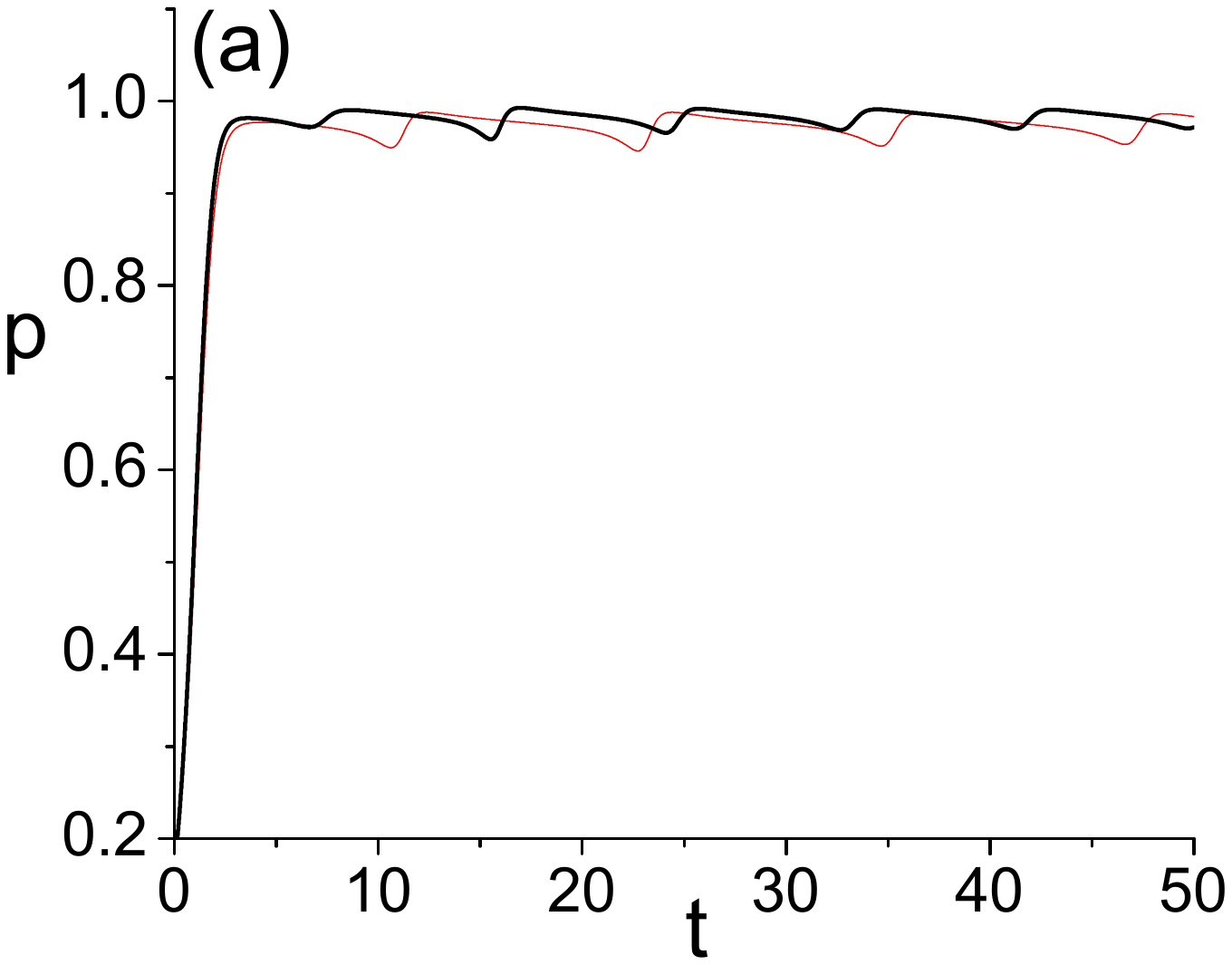} 
	\includegraphics[scale=0.29]{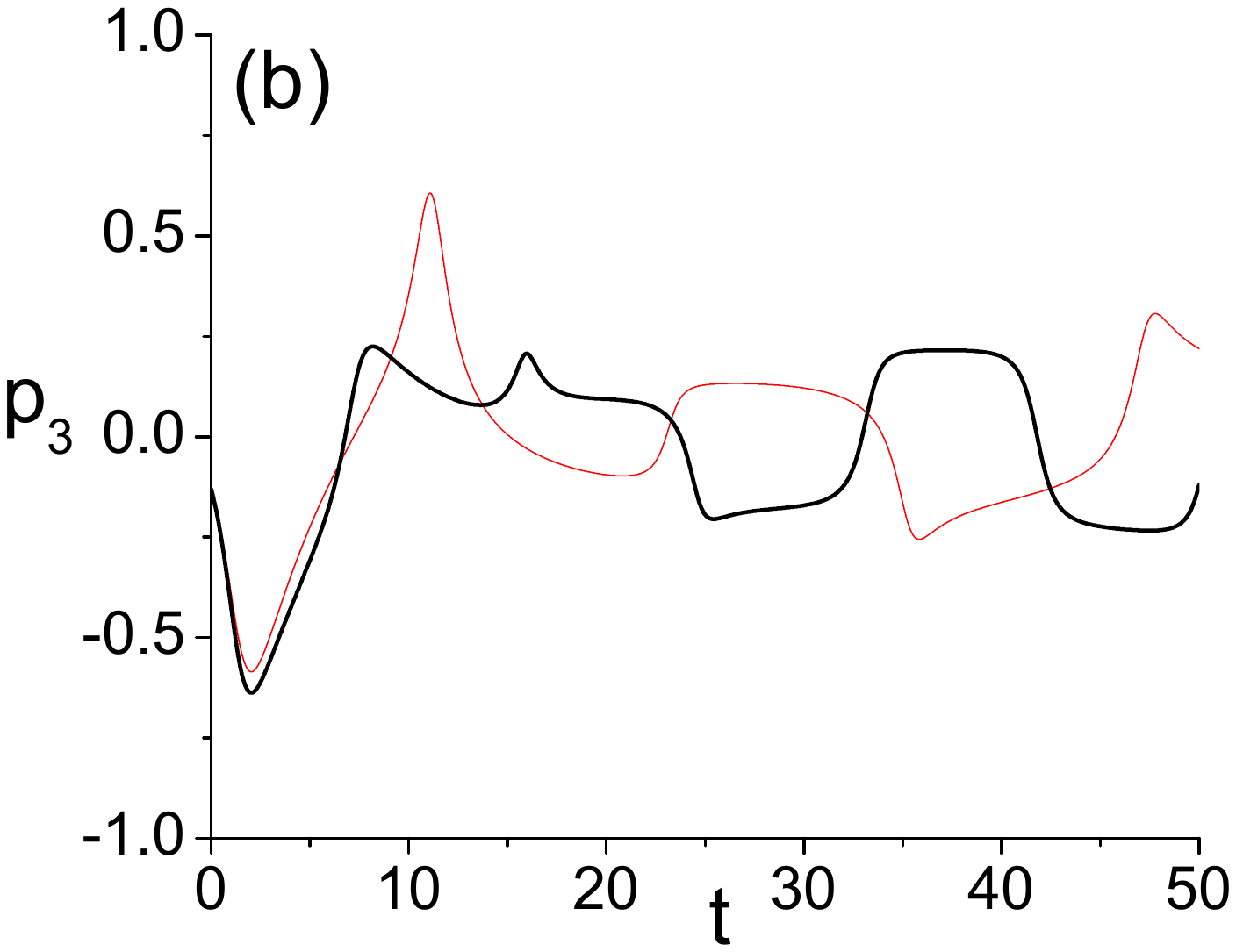} 
	\includegraphics[scale=0.29]{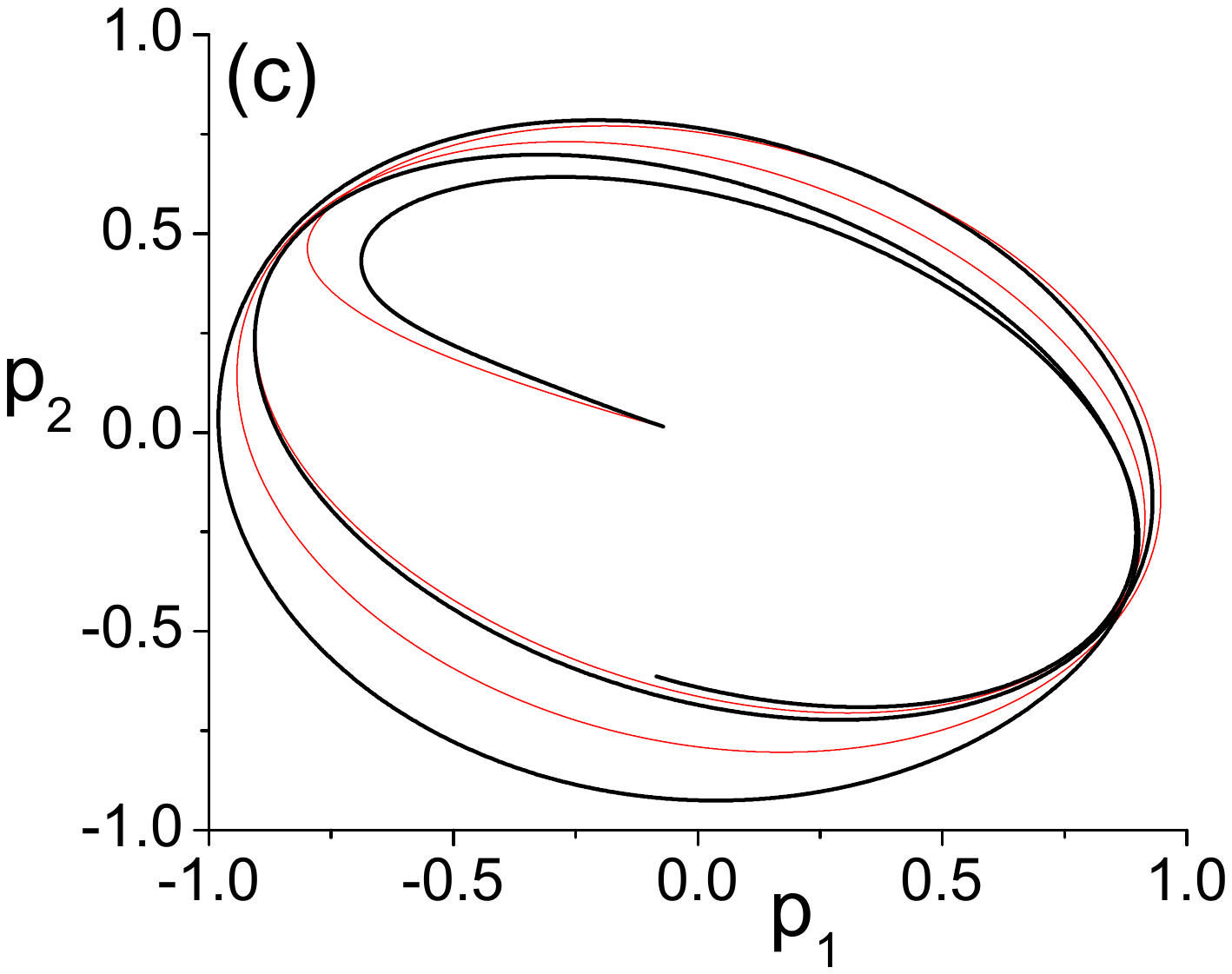} 
	\includegraphics[scale=0.29]{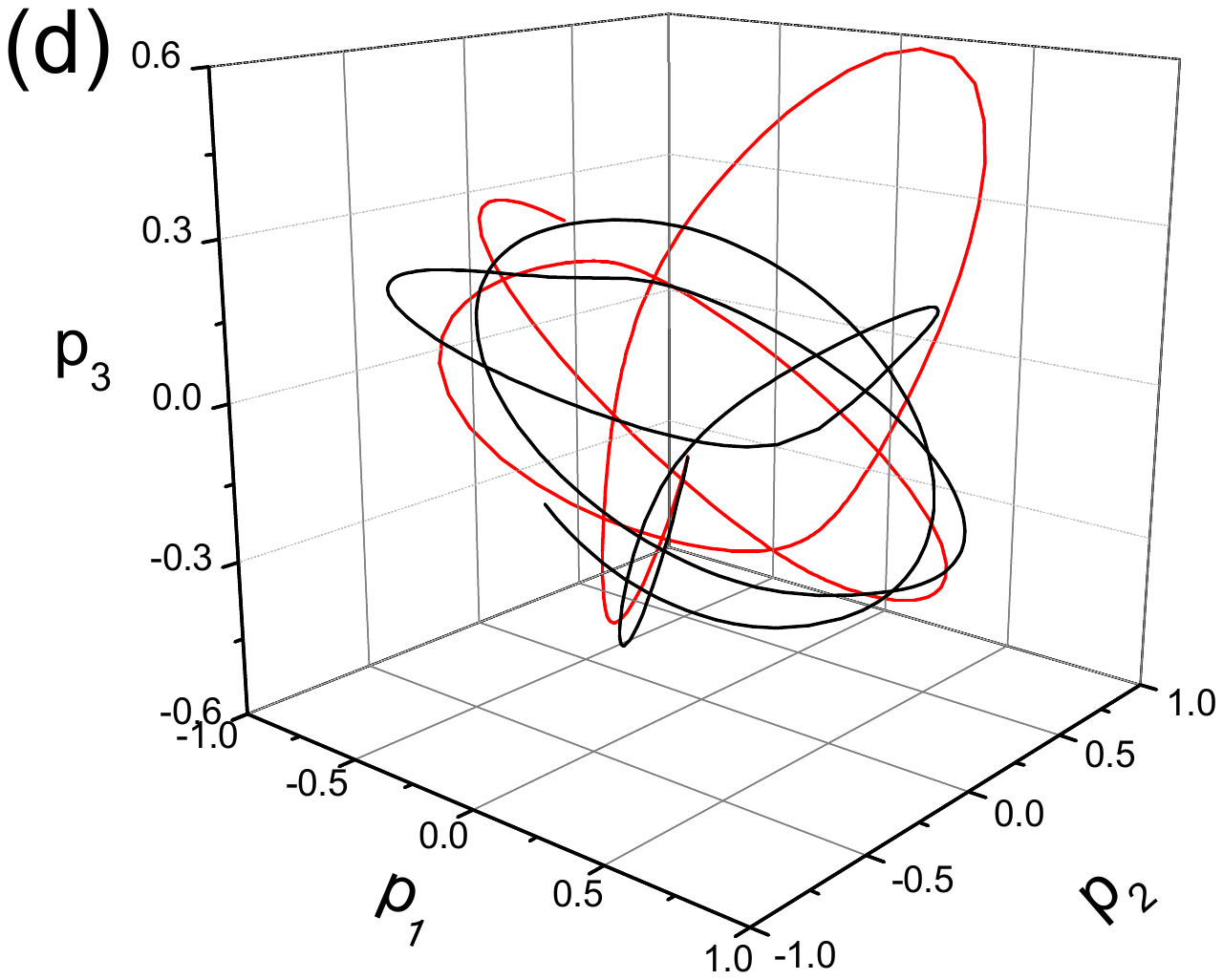} 
	\caption{Evolution of order parameter for the coupling matrix (\ref{4dmat}). Module of the order paramter (a) and $p_3$ (b) as a function
		ot time. Panel (c) shows the $p_1 \times p_2$ projection and (d) the $p_1 \times p_2 \times p_3$ projection. In all plots the red thin line
		shows the results with Euler's method and the thick black line the results using Eq.(\ref{evol4d}). }
	\label{fig3}
\end{figure}

Next I consider $N=100$ particles interacting with coupling matrix
\begin{equation}
	\mathbf{K} = \left(
	\begin{array}{cccc}
		k_1 \cos \alpha & k_1 \sin \alpha & 0 & 0\\
		-k_2 \sin \alpha & k_2 \cos \alpha & 0 & 0\\
		0 & 0 & k \cos \beta & k \sin \beta\\
		0 & 0 & -k \sin\beta & k \cos \beta 
	\end{array}
	\right),
	\label{4dmat}
\end{equation}
with $k_1=2.6$, $k_2=1$, $k=1.5$, $\alpha=0.5$ and $\beta=0.2$. I consider the full matrix of natural frequencies
\begin{equation}
	\mathbf{W}_i = \left( 
	\begin{array}{cccc}
		0 & -\omega_{6i}  &  \omega_{5i} & -\omega_{4i} \\
		\omega_{6i} & 0  &  -\omega_{3i}  & \omega_{2i}\\
		-\omega_{5i} & \omega_{3i} & 0 & -\omega_{1i}  \\
		\omega_{4i} & -\omega_{2i}  & \omega_{1i} & 0\
	\end{array}
	\right)
	\label{wmat4}
\end{equation}
and choose all $\omega_{ij}$ Gaussian distributed with zero average and width $\delta=0.2$. 
Fig.\ref{fig3} shows comparisons between simulations with Euler's method and Eq.(\ref{evol4d}). The
overall effect on the module of $\vec{p}$ is not so large, but the trajectory $\vec{p}(t)$ changes considerably
as small errors propagate after $t \approx 10$. Simulations using RK4 method also works well and, for this range
of time, falls on top of the curve generated with Eq.(\ref{evol4d}.

\section{Conclusions}

Lee and collaborators have recently shown that the application of Euler's method to the integration of the 3D Kuramoto
model results in numerical errors that cannot be corrected by decreasing the simulation time steps \cite{lee2023improved}. The sources
of error are the change in norm of the propagated vectors, that move away from the sphere and need to be manually projected back onto its surface, and the intrinsic error in the direction of the rotation axis produced by Euler's procedure. 
They have shown that proper integration must consider rotation of the particles over the sphere and they do that with a sequence of rotations 
using Euler's angles. Here I proposed a slightly different way to perform such rotations that might be easier to implement 
numerically and that can be extended to higher dimensions. The method consists in writing the dynamical equations in the form
$\dot{\vec{\sigma}}_i = e^{V_i} \vec{\sigma}_i$ and use Cayley-Hamilton's theorem to write $e^{V_i} = \sum_{k=0}^{D-1} \beta_k V_i^k$, 
a finite sum that contains all orders of the exponential. I provided explicit formulas for the cases of $D=2$, 3 and 4 
and compared the results with different integration methods. I have also shown, numerically, that the fourth order Runge-Kutta 
method does work well in three and four dimensions. The errors generated by projecting the vectors back to the sphere after each 
integration time step do not seem affect the accuracy of results in these cases.

It is not clear if the form of Eq.(\ref{kuramotogenkh}) is useful for exploring other properties of the Kuramoto dynamics or if
it is just a mathematical trick to facilitate numerical integration as I have done here. As a final remark I note that the same formalism can be 
used to integrate the Kuramoto model in the presence of external forces. Adding a period force to each oscillator like \cite{Childs2008}
\begin{equation}
	\dot{\theta}_i = \omega_i + \frac{k}{N} \sum_{j=1}^N \sin{(\theta_j-\theta_i)} + F \sin(\theta_i-\Omega t)
	\label{kuramotof}
\end{equation}
results in the vector equation
\begin{equation}
	\frac{d \vec{\sigma_i}}{d t} = \mathbf{W}_i \vec{\sigma_i} +  [(k\vec{p} - \vec{F}) - (\vec{\sigma}_i \cdot (k\vec{p} - \vec{F})) \vec{\sigma_i}]
	\label{eq3conc}
\end{equation}
where $\vec{F} = (F\cos(\Omega t), F\sin(\Omega t))$ \cite{barioni2021complexity}. This amounts to replace $k \vec{p}\,$ by 
$k\vec{p} - \vec{F}$ and can be extended to multi-dimensional systems with matrix coupling as 
\begin{equation}
	\frac{d \vec{\sigma_i}}{d t} = \mathbf{W}_i \vec{\sigma_i} +  [(\mathbf{K}\vec{p} - \vec{F}) - (\vec{\sigma}_i \cdot (\mathbf{K}\vec{p} - \vec{F})) \vec{\sigma_i}]
	\label{eq4conc}
\end{equation}
where $\vec{F}$ is the generalized D-dimensional force applied to the oscillators. 
Therefore, to take into account external forces, the only change in the integration algorithm is to replace the vector $\vec{p}_K(t) = \mathbf{K} \vec{p}(t)$ 
in Eq.(\ref{pk}) by $\vec{p}_K(t) = \mathbf{K} \vec{p}(t) - \vec{F}(t)$.

\begin{acknowledgments}
	It is a pleasure to thank Joao U.F. Lizarraga for helpful comments. This work was partly supported by FAPESP, grant 2021/14335-0 (ICTP‐SAIFR) and CNPq, grant 301082/2019‐7. 
\end{acknowledgments}

\clearpage 
\newpage
\bibliographystyle{ieeetr}

\end{document}